\documentclass[11pt,a4paper]{article}

\usepackage{xspace,graphicx,url}

\newcommand{\gaia}{\textsf{GAIA}\xspace}
\newcommand{\ieeeZ}{IEEE 802.15.4\xspace}

\newcommand{\wifi}{WiFi\xspace}
\newcommand{\myparagraph}{\smallskip\noindent\textbf}

\begin{document}

\title{IoT-based Big Data Analysis of School Buildings Performance}

\author{Ioannis Chatzigiannakis\thanks{Sapienza University of Rome, Italy}, 
Georgios Mylonas\thanks{Computer Technology Institute \& Press (CTI), Greece},\\ 
Irene Mavrommati\thanks{Hellenic Open University, Greece} and 
Dimitrios Amaxilatis\thanks{Computer Technology Institute \& Press (CTI) and University of Patras, Greece}}

\maketitle

\begin{abstract}
The utilization of IoT in the educational domain so far has trailed other more commercial application domains. In this chapter, we study a number of aspects that are based on big data produced by a large-scale infrastructure deployed inside a fleet of educational buildings in Europe. We discuss how this infrastructure essentially enables a set of different applications, complemented by a detailed discussion regarding both performance aspects of the implementation of this IoT platform, as well as results that provide insights to its actual application in real life, both from an educational and business standpoint. 
\end{abstract}

\section{Introduction}
\label{sec:intro}
Wireless Sensor Networks have been the starting point for a tremendous development that has gradually led to the realization of the Internet of Things (IoT). Today, there is a large variety of hardware and software to choose from that is easy to set up and use in an increasing set of real-world application domains~\cite{Chatzigiannakis2011103}. One such important domain is education: the deployment of a variety of sensors (e.g., for monitoring electricity consumption, environmental conditions, be them indoor or outdoor) across school buildings can produce real-world data that could be directly used in educational activities or provide input to business processes, making financial sense in terms of cost savings.

Another thing to consider is climate change and our response as a society through the transfer of green technologies inside schools. The last few years an emphasis has been given on environmental awareness via education. In many cases, this is achieved by employing lab activities based on off-the-shelf IoT sensors, planned specifically for students as part of science classes. Such hardware is in many aspects identical to what much of the IoT research community is currently using; thus, the potential in combining recent results coming from the research community, with the educational activities already in the curricula of many schools is  promising. 

In this context, this chapter deals with energy and environmental awareness as a part of school educational activities. This is handled in two ways: a) by addressing energy footprint and energy consumption, via individual and group class activities, by using IoT sensors and gamification elements using real sensor data from familiar environments, and recording changes in behavior that affect directly energy consumption, and b) To raise environmental awareness of the systemic nature of changes, (affecting sustainability of ecosystems, climate, etc) via a quest for inquiry and knowledge using data from sensors distributed across the European continent. 

Several studies document the ability of students to influence choices made by their families related to environmental issues~\cite{schelly12}. The research interviews conducted in~\cite{powerdown} made clear that energy conservation insights learned in school can be applied at home by students and their families. Since about 27\% of EU households include at least one child under the age of 18~\cite{eurostat10}, targeted efforts of reaching families of children and young people will scale further to reach a large portion of the EU population and multiply the benefits towards sustainability of the planet.

Having the above in mind, the aim is to provide a system for monitoring a fleet of educational buildings, enabling a number of different application and implementation aspects. In short, our system aims at the following properties:

\begin{itemize}
	\item \textit{openness}, supporting a number of different IoT ecosystems,
	\item \textit{versatility}, supporting different application domains, e.g., energy efficiency and educational scenarios,
	\item \textit{scalability}, supporting a very large number of buildings and IoT sensing endpoints,
	\item \textit{up-to-date support of modern practices} in the design of the system, i.e., cloud-based solutions, easy deployment, etc.
\end{itemize}

It is currently being used in 18 educational buildings in Europe in different countries, in the context of an EU-funded research project, called Green Awareness In Action (GAIA\cite{gaia-website}).The system utilizes a number of different IoT technologies as infrastructure installed inside these school buildings (utilizing up to 850 sensors), while it follows an open, cloud-based approach for its implementation, which enables the development of applications on top of it. In its current setup, this deployment produces daily close to 400MB of real-world sensor data, resulting in a yearly data volume of approximately 140GB. Depending on the actual application domain in which such data are used, there is a lot of variance on the actual requirements for data granularity, i.e., sampling rate for generating data. For obtaining near real-time information on the building status, it is necessary to have a small sampling period, even in the scale of milliseconds, e.g., in the case of energy disaggregation. Similarly, usually an averaging scheme is used so that data storage and processing requirements are kept at low level, since more data-intensive aggregation methodologies increase the requirements for processing resources.

In this context, such a large-scale IoT infrastructure deployed at a large number of public and private buildings for a long period of time will generate, handle, transfer and store a tremendous amount of data, which cannot be processed in an efficient manner using current platforms and techniques. Since the community is gradually moving from vertical single-purpose solutions to multi-purpose collaborative applications interacting across industry verticals, organizations and people, a cloud-based approach is utilized. The necessity for data collection, storage and availability across large areas, the demand for uninterrupted services even with intermittent cloud connectivity and resource constrained devices, along with the necessity of sometimes near-real-time data processing in an optimal manner, create a set of challenges where only holistic solutions apply. IoT platform targeting large scale deployments are expected to rapidly process the constantly accumulated data in the edge. 

The remaining chapter is structured as follows: in Section~\ref{sec:sota} previous and related work is presented briefly. The benefits of using data collected from IoT deployments in education is presented in Section~\ref{sec:stem}. In Section~\ref{sec:iot} the high-level design of an IoT platform targeting education activities is presented addressing the end-user requirements. Having in mind the design goals and end-user needs, in Section~\ref{sec:gaia} the \gaia platform is presented, one among the very few IoT systems that have been developed with a focus on education. Specific implementation decisions are presented in detail and indicators on the performance of the IoT platform are provided. In Section~\ref{sec:data} provides an analysis of the data collected from the \gaia platform in terms of the performance of the educational buildings. We conclude in Section~\ref{sec:conclusion}.

\section{Related Work}
\label{sec:sota}
The approach of promoting sustainable behavioural change through activities targeting the public building sector falls within the scope of several research projects. Recent examples like \cite{s18020537} and \cite{s17092054}, focus on a variety of public buildings utilizing an IoT infrastructure over which applications like gamified experiences or blockchain-based transactions promote behavioral change among the occupants of such buildings. More focused on school buildings were the Veryschool \cite{veryschool} and Zemeds \cite{zemeds2015} projects, producing recommendation and optimization software components, or methodologies and tools. \cite{school-future} produced several guidelines and results regarding good energy saving practices in an educational setting. There is however a general lack of works focusing exclusively on the application of the IoT paradigm in an educational realm, in combination with large-scale real-world data. This exact aspect is examined in our work, aiming to combine energy savings and educational goals at the same time.

Overall, the proliferation of relatively cheap IoT hardware has led several parties to establish large-scale building infrastructures, leading to large datasets created and the inevitable exploration of a number of potential utilization of such data inside applications. \cite{building-data-genome} discusses an attempt to create a big data dataset for building data with respect to energy, while \cite{big-building-data} presents an approach towards systems built over the use of big data from buildings. \cite{ml-cooling} and \cite{batra2015} are 2 recent examples describing ways to utilise big data generated inside buildings in order to implement applications towards better thermal comfort and energy savings respectively.

Certain aspects of the system discussed here, place users in the monitoring loop, as a first step towards raising awareness. This aspect, categorized as crowdsensing or participatory sensing \cite{Burke06participatorysensing} in which users collect relevant data for  applications such as urban planning, public health, creative expression, etc., aims at emphasizing the personalization factor of such systems, among other aspects. This approach has been employed by the  Cornell Laboratory of Ornithology \cite{doi:10.1080/09500690500069483} in a science educational project on bird biology, while in \cite{6296848} the authors describe trials for air quality, water quality and plant disease monitoring. Similarly to our context, \cite{6990342} presents a solution combining a deployed and participatory sensing system for environmental monitoring. In \cite{Heggen:2013:PSR:2405716.2405722} the authors discuss the value of participating to project like these for students, concluding that ``Students are gaining deep domain-specific knowledge through their citizen science campaign, as well as broad general STEM knowledge through data-collection best practices, data analysis, scientific methods, and other areas specific to their project''.

In the past several approaches have been proposed in order to address the potentially huge number of sensor data arriving from the IoT domain, each one of them applied in different parts of the network architecture~\cite{Chatzigiannakis2007466,Chatzigiannakis2005376,DBLP:conf/iot/ChatzigiannakisHKKKLPRT12,5633643}. Starting from the low-end devices, the approach of in-network aggregation and data management has been proposed where sensor devices follow local coordination schemes in order to combine data coming from different sources and/or within the same time period based on similarities identified using data analysis. Usually these techniques operate in combination with network-level routing protocols and/or lower-level medium access control protocols~\cite{DBLP:conf/anss/ChatzigiannakisKN05}. For an overview of different techniques and existing protocols see~\cite{Fasolo:2007:IAT:2212993.2213367}. Since this approach relies on spatial and temporal correlation without taking into consideration semantic correlation of the data, very few theoretical algorithms are used in real-world deployment since they significantly limit the concurent support of different high-level applications.

Information on the design aspects of the \gaia IoT platform presented in this chapter are available in~~\cite{s17102296}. Technical details for efficient stream processing of IoT data collected from school buildings utilizing edge resources are provided in~\cite{akrivopoulos-etfa-2017}. The design of open-source hardware end-devices for monitoring energy consumption and environmental conditions inside schools are presented in~\cite{Pocero2017}. The use of IoT data generated inside school buildings for addressing behavioral change towards energy efficiency is discussed in~\cite{gaia-workshop-geneva}. The use of the open IoT platform for the development of a building management application is presented in~\cite{synelixis-paper}.

\section{IoT and Real-world Data in Education}
\label{sec:stem}
One approach for addressing the climate change problem is through the development and transfer of green technologies. In the context of reducing the energy spent in residential buildings, new technologies have been introduced improving the energy efficiency of buildings. In fact, till now the dominant approach was to use energy-efficient infrastructure and materials to reduce energy consumption of buildings. However, the rates of construction of new buildings as well as the rates of renovation of existing buildings are both generally very low~\cite{bpie11} to expect a significant effect on the total amount of energy spent in our everyday life at a global level. Similarly, the approach for reducing the energy consumption in transportation focuses on improving the energy efficiency of motor engines. Also here, given the rate of change of existing fleets with energy efficient one, it is very challenging to save energy in this sector through this approach~\cite{iccs08}. 

An alternative approach, that has recently gathered attention, is the promotion of energy consumption awareness and behavioral change on people. The main concept is that to address  global climate change, people should be informed regarding mitigation actions and sustainable behaviors. In other words, it requires a change in citizens' behavior and practices~\cite{eea13}. Reports indicate that citizens making efficient use of energy in their everyday life can lead to large energy and financial savings, as well as potentially to a substantially positive environmental impact~\cite{eea13}. 

Raising awareness among young people and changing their behavior and habits concerning energy usage is key to achieving sustained energy reductions. At EU level, people aged under 30 represent about a third of the total population \cite{eurostat12}. Thus, by targeting this group of citizens, a larger part of the population is also affected. Additionally, young people are very sensitive to the protection of the environment so raising awareness among children is much easier than other groups of citizens (e.g., attempts made to achieve behavioral change and establish new environment-friendly habits to children regarding recycling have had high success rates).


We argue that it is essential to operate an IoT platform that will on the one hand facilitate monitoring and profiling energy use of users and buildings and on the other hand will provide guidelines and recommendations for better energy management by users. Such a platform will increase the self-awareness of users regarding their energy use profile and by proper and continual recommendations will stimulate their behavior change toward more energy economical activities and habits. 


\subsection{End-user Requirements}

A main objective of environmental sustainability education and energy efficiency awareness is to make students aware that energy consumption is largely influenced by the sum of individual behaviours (at home, school, etc.) and that simple behaviour changes and interventions in the building (e.g., replacing old lamps with energy-efficient ones) can have a great impact on achieving energy savings. IoT technologies can support these initiatives by mediating people's interaction with the environment in order to provide immediate feedback and actually measures the impact of human actions while automating the implementation of energy savings policy and at the same time maintaining the comfort level perceived by people. 

From the end-user point of view, the system aims to support different end-user groups. Within school buildings, there are several such groups: students, educators, building administrators. There is thus a need for the system's interface to provide services and information in a way that suits all of these different end-users. In some cases, unifying different school buildings into a single view is necessary to make interaction simpler, make data visualisation more natural and create an environment that conveys valuable insights and clear actions related to general as well as specific aspects of the participating building ecosystem. In other cases, the educational scenario requires differentiating each building or room within the building to stress the unique features and emphasize on the needs of each individual involved.

Teachers could potentially use collected data and analytics during class to explain to pupils basic phenomena related to the parameters monitored and organize student projects, where each student monitors specific environmental parameters at their home. In addition, collected IoT data could feed applications informing building managers about the energy profile performance of the building and specific equipment. Similarly, the IoT data collected in schools can be made available to the scientific community, so that studies can be performed on a common dataset and results can be more easily compared. 
Monitoring school buildings situated in different countries can help, e.g., to identify usage or energy consumption patterns. This, in turn, can be utilized to make comparisons or realize competitions through social networking and game applications (e.g., students of school A compete with students of school B in answering energy awareness questions). The availability of actual measurements of environmental parameters, such as energy consumption, indoor and outdoor luminosity, temperature, noise, and so on, also enables the realization of a number of diverse education-related applications and scenarios. For example, student engagement could be fostered via projects where students monitor environmental parameters at their class or home, or programming software using the data provided by the platform utilizing the available APIs. 

\subsection{IoT Platform Design Aspects}

In this section the most fundamental goals that an IoT-based platform for real-time monitoring and management of public school buildings are presented. 

The system's first goal is to build upon an IoT infrastructure that (i) spans throughout a big number of buildings and (b) spreads through several rooms, incorporating various kinds of sensors. Providing a uniform set of hardware devices for such a large number of building is very difficult, if not impossible. We expect the necessary IoT infrastructure to be deployed in phases - in each phase the IoT infrastructure will be extended to cover additional buildings or introduce new sensors. It is therefore crucial to follow an open standards approach throughout such a system; use open source software, protocols \textit{and} hardware components, in order to maximise the adaptability and reusability of the system. 

Another important goal of a system deployed at such a scale and operating in real-time is the need for quickly handling, storing and analyzing vast quantities of data collected from the IoT nodes. Providing an efficient software platform for data management will enable direct comparisons of energy efficiency between different buildings and cities, taking into account the environmental parameters as well, e.g., rain or low temperatures, etc., helping users quantify their actions in terms of the respective impact to the environment.

Lastly, the system should be expandable and easy to interface with other systems or components. 
As examples, implementing a basic quiz-based approach to support school classes on sustainability or a serious game-based approach to learning, are ideas that can be augmented by using real-time data from our system. In terms of administration of an institutional infrastructure, interfacing with commercial products and new cloud-based services coming out of the research labs is absolutely necessary to make sure that the IoT infrastructure will not become obsolete.

\section{Design aspects of an IoT Platform targeting Education activities}
\label{sec:iot}
An IoT platform targeting educational activities should be based on the principle that continuous monitoring of the progress of students positively contributes towards reducing the energy consumption and successful behavior change. Since the IoT deployment is multi-site and multi-country can motivate, for example, to identify energy consumption patterns in different countries and across different climate zones. This can be used to make comparisons or competitions; for instance, students of school A compete with students of school B in efficiency. This could also help understanding cultural differences with respect to energy efficiency awareness and sustainability.

Such an IoT platform needs to be designed to enable the easy and fast implementation of applications that utilize an Internet-of-Things infrastructure. It should offer high scalability both in terms of users, number of connected devices and volume of data processed. The platform accommodates real-time processing of information collected from mobile sensors and smart phones and offers fast analytic services. The platform architecture is organized in three layers:

\myparagraph{End-Device Level:} consists of IoT Metering devices (electricity, heating, environment) deployed in educational buildings feed the system information relating to their energy consumption and the indoor/outdoor environmental conditions.

\myparagraph{IT Service Ecosystem Level:} comprises the overall system with the capacity to store, manage, analyze and visualize data regarding the behavior of users, the ambient conditions of the buildings and the energy consumption. This level consists of a set of services, exposed through a specially designed set of APIs for: data acquiring, storage and diffusion, user and building modeling/profiling, data visualization, and user recommendation. On top of these services, \gaia offers end-user applications to engage users.

\myparagraph{User Involvement Level:} consists of end-user applications for providing data analytics and recommendations to the targeted user groups, social networking applications and an educational game for students.

\subsection{End-device Level}

\begin{figure}[!t]%
\includegraphics[width=\textwidth]{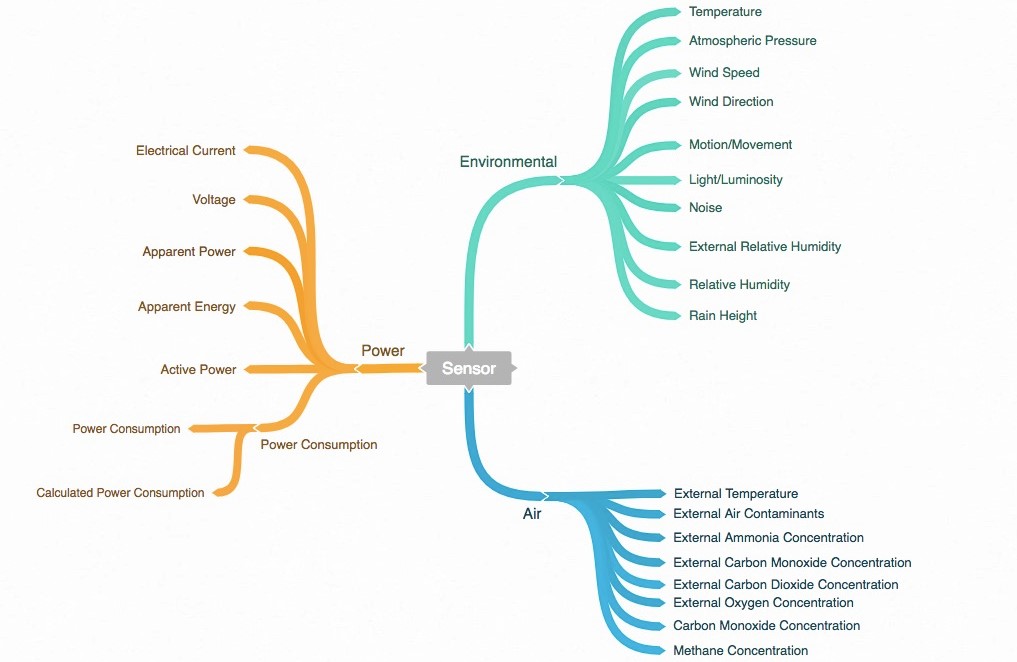}
\caption{\label{sensor-types}The sensing capabilities of an IoT platform targeting School Buildings}
\end{figure}

A diverse set of deployed devices constitute the End-device level organized in four main categories: (1) classroom environmental comfort sensors (devices within classrooms); (2) atmospheric sensors (devices positioned outdoors); (3) weather stations (devices positioned on rooftops); and (4) power consumption meters (devices attached to the main breakout box of the buildings, measuring energy consumption). For a graphical representation of the different sensing types used throughout the end-device level, see Fig.~\ref{sensor-types}. 

The end-devices that are deployed indoors form wireless networks (802.15.4 or WiFi) and communicate with their respective edge devices by establishing ad hoc multihop bidirectional trees, setup in the time of the deployment and maintained throughout the network lifetime. The end-devices deployed outdoors are connected either via Power Over Ethernet cables to transfer both electricity and maintain communication over a single cable, or they have a WiFi network connection and are supplied with batteries and solar panels to harvest energy from the sun. On the transport and session layers, the devices communicate either using a custom protocol or Zigbee for the discovery of resources and transmission of measurements. 

\myparagraph{Open-design Power Consumption meters} are installed to measure the apparent power and average power consumption of a school building. Meters are situated on the general electricity distribution board of each building to measure each one of the 3-phase power supply of the building. These devices are Arduino-based and use XBee modules in order to access the \ieeeZ network. For more details regarding the design and technical specification of the devices see~\cite{gaia-hardware}.

\myparagraph{Meazon Power Consumption meters} provide an efficient and cost effective way to monitor energy wirelessly in real-time. The devices provide advanced meter points for current, voltage, frequency, active and reactive power/energy. These sensors communicate over \ieeeZ networks with a proprietary gateway device that is either connected to the Internet via Ethernet or use 3G in order to communicate directly with Meazon's proprietary cloud services.

\myparagraph{Open-design Environmental Comfort meters} measure various aspects affecting the well-being of the building's inhabitants, such as thermal (satisfaction with surrounding thermal conditions), visual (perception of available light) comfort and overall noise exposure. They also monitor room occupancy using passive infrared sensors (PIR). These devices are also Arduino-based and use XBee modules in order to access the \ieeeZ network. For more details regarding the design and technical specification of the devices see~\cite{gaia-hardware}.

\myparagraph{Open-design Weather and Atmosphere Stations} provide information on the outdoor atmospheric conditions including precipitation levels, wind speed, and direction. The \textit{atmospheric meters} monitor atmospheric pressure and concentration of selected pollutants, to provide insights on the pollution levels near school buildings. These devices are also Arduino-based, are connected either via Ethernet or via \wifi and are powered using Power-Over-Ethernet, or are plugged into the sockets of the building when available. For more details regarding the design and technical specification of the devices see~\cite{gaia-hardware}.

\myparagraph{Synelixis Weather station} are off-the-self devices that use \wifi connectivity to avoid installing additional cables on the roofs of the buildings, as well as energy harvesting via solar panels. The end-devices communicate via \wifi directly to proprietary cloud services.

\subsection{IT Service Ecosystem Level}

A key principle is to foster education and experimentation activities leveraging the availability of actual measurements of environmental parameters, such as energy consumption, indoor and outdoor luminosity, temperature, noise, etc. The IT Service ecosystem allows to feed end-user applications with such monitoring data in order to enhance educational processes with data from real-life scenarios. 
The system hides the heterogeneity of sensors deployment to end-user applications by managing the communication with the sensor infrastructures through dedicated software modules, handling and storing data according to a uniform model and offering such data to applications through uniform interfaces. When a new school is added to the system, existing applications can easily access sensors’ data with only some minor reconfiguration. Moreover, measurements provided by sensors may need some processing to become meaningful and usable by end-user applications. Therefore, filtering, reduction, and formatting operations are performed on raw data flows acquired by networks of sensors to transform them in information that is suitable for storage and higher level processing. More in detail, formatting operations are applied to convert heterogeneous data obtained from different sensors into a uniform naming convention. In this way, applications and higher-level processing services can seamlessly access data gathered from different sources.

The IT Service ecosystem provides a set of data storage and processing services for extracting, persisting and disseminating information related to energy-efficiency and resource optimization that are derived from measurements collected from sensors or by end-users through participatory sensing. The storage service provided is the key to transform data in motion (i.e., data flows generated by sensors) to data at rest, i.e., data from readily accessible storage facilities~\cite{cisco-iot-paper}. Indeed, it is widely recognized that in the IoT domain, most end-user applications need to access data on a non-real time basis~\cite{Stallings}.

The data storage service offers APIs to allow end-users (using  applications that leverage such APIs) to access, provide and manage information regarding the profile, the topology and other relevant information characterizing the school buildings and their usage (e.g., geographical location, organization of spaces, number of students, opening and closing times).

These data can then be accessed by modules that perform high-level processing tasks (i.e., analytics and recommendation engine) for extracting further meaningful and useful information that could help users (especially building managers) in:
\begin{itemize}

\item visualizing and understanding the energy-consumption profile of a building;
\item being aware of possible anomalies in energy consumption;
\item receiving suggestions for actions and behavior-based changes possibly leading to energy reductions while maintaining the comfort of the learning environment;
\item managing and customizing suggestions notified at the occurrence of events of interests related to energy consumption (e.g., list of equipment to be checked and eventually turned off before holiday periods).

\end{itemize}

In addition, the IT Service ecosystem also provides support to participatory sensing approaches~\cite{estrin:predictions} by offering APIs allowing end-users applications to manually provide consumption data (e.g., via csv and MS excel files) or web forms.

Finally, the IT Service ecosystem also provide end-users applications with authentication and authorization services, leveraging existing standards and offering basic profile storage services. Indeed, GAIA end-user applications need to differentiate permissions according to user’s roles (e.g., student, teacher and building manager).

\subsection{User Involvement Level}

Including the users in the loop of monitoring their daily energy consumption is a first step towards raising awareness. In an educational environment, this step can be further enhanced and capitalized in the framework of educational activities with the support of the application offered at the User Involvement Level. In essence, the application set complements an educational approach that encourages customization to the specific requirements of each school, where each school can ``fine-tune'' which tools are used, when and for how long, during a school year. Overall, the educational activities in each school are based on data produced within the respective buildings, while the effects of changing certain behaviors can be detected and quantified. 

The system offers a rich set of APIs to access the information of the deployed sensors as well as the data collected in real time. These APIs have been used to develop multiple applications that are designed to help the building managers, teachers and students in their day-to-day activities in the school buildings (educational or not). Among others, third parties have developed a building management application (BMS)~\cite{synelixis-paper}, a participatory sensing application, as well as a set of in class activities accompanied by sensing and visualization tools that help students better understand their environment and the natural effects monitored.

\myparagraph{Educational Building Management system} is a multi-school BMS developed using the API exposed by the IT Service ecosystem. Building managers are able to inspect real-time energy usage, see results from a comparison with similar buildings or within the same building during different time periods (e.g., previous years), and receive energy efficiency recommendations. 

\myparagraph{Participatory Sensing application} offers a way to complement the existing IoT infrastructure (End-device Level) and provide additional data from the school buildings monitored by the system. Smartphones and tablets are used to provide additional readings from inside the school building, e.g., luminosity or noise levels, while participants can also manually enter data such as electricity meter readings. The teacher can initiate participatory sensing sessions during the courses from the main portal of the project and then students can use phones and tablets to gather data in real time and then review them in class.

\myparagraph{In-class Activities and Gamification applications} with an educational focus have been implemented to support educational content based on the data produced by the IoT infrastructure inside the monitored school buildings. These are based on the fact that students are more driven to engage in class activities regarding sustainability and energy efficiency when the data utilized originates from their environment and are near real-time. The activities are built around a sensor kit, which the students use to build a small interactive installation that visualizes environmental and energy consumption data from school classrooms. 

\section{The \gaia IoT Platform}
\label{sec:gaia}
The \gaia real-world IoT deployment is spread in 3 countries (Greece, Italy, Sweden), monitoring in real-time 18 school buildings in terms of electricity consumption and indoor and outdoor environmental conditions. Given the diverse building characteristics and usage requirements, the deployments vary from school to school (e.g., in number of sensors, hardware manufacturer, networking technology, communication protocols for delivering sensor data, etc.). The IoT devices used are either open-design IoT nodes (based on the Arduino popular electronics prototyping platform, see~\cite{Pocero2017}) or off-the-shelf products acquired from IoT device manufacturers. The data collected is used as part of series of educational scenarios whose goal is to educate, influence and attempt to transform the behavior of elementary school students through a series of trials conducted in the educational environment and in homes. Feedback mechanisms notify the students on current energy consumption at school and in this way assist towards raising awareness regarding environmental effects of energy spending and promote energy literacy by educating the users.

The cloud services offer real time processing and analysis of unlimited IoT data streams with minimal delay and processing costs. Storage services use state of the art solutions like NoSQL and time series databases to ensure maximum scalability and minimal response times. In more detail, the cloud services deliver a set of services that are critical for all IoT installations:

\myparagraph{Continuous computation engine} provides real-time, fast, and reliable processing of data collected from IoT devices, smart phones and web-services. The computation engine is capable of processing a large amount of data collected from sensor nodes within just seconds. 

\myparagraph{Online analytics engine} operates on the data produced by the continuous processing engine. The post-processing of the data is done to support business intelligence. The online analytics engine allows to organize large volumes of data and visualize them from different points of view.

\myparagraph{End-to-end security} is established across the components of the architecture. All supported services are compliant with the current standards for Internet security. Communication throughout the service infrastructure is encrypted using data encryption standards like AES and TLS/SSL technologies.

\myparagraph{Access management and Authorization} is managed in real-time down to specific user, device or time of day.

\myparagraph{Data Storage \& Replay services} are responsible for the persistence of all data entering the system in their original format and associated with the output of the continuous processing engine and online analytics engine. Data streams can be forwarded at a later time to different components. Offline processing of data is facilitated for archiving services or for bench-marking different versions of components.

The services described above can be accessed via a well-defined set of APIs. The Data API is comprised of Real-time Data API and Historical Data API. Historical Data API allows retrieval of historical data registered into the platform by any device and also aggregated summaries (maximum values, minimum values, average values). Real-time Data API is a streaming API which gives low latency access to new Data registered to the platform. Directory API and AA (Authentication/Authorization) API describe how to create and manage devices, users and authorization roles.

\begin{figure*}[!t]
\centering
\includegraphics[width=0.9\textwidth]{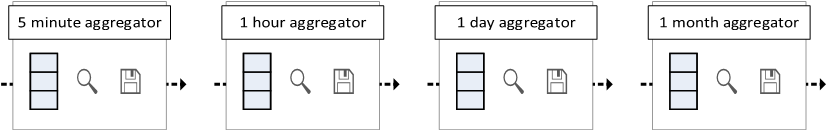}
\caption{\label{fig-sparks-chain} Analytics process chain}
\end{figure*}

\subsection{Continuous Computation Engine}
\label{sec:cce}

The Continuous Computation Engine is a central part of the system architecture, and is responsible for the timely processing of the streaming data arriving from the sensors deployed across all the buildings. It is composed by a \textit{Message Bus} receiving all the messages arriving from the sensors, the  \textit{Process Engine} which provides the analytics and the \textit{Storage System} which is used for storing those results. The Process Engine receives events from multiple sensors and executes aggregate operations on these events. The output of the engine is stored at the \textit{Storage System}.

Sensors and actuators produce (periodically or asynchronously) events that are sent to the \textit{Continuous Computation Engine} via a message bus. Those events are usually tuples of pairs: value and timestamp. All data received are collected and forwarded to a specific queue where they get processed in real time by the \textit{Process Engine}. The \textit{Process Engine} supports a number of processing topologies based on the data type of the sensor. Each processing topology is responsible for a unique type of sensor such as general measurement sensors (temperature, wind speed, etc.), actuators, power measurement sensors, etc. The produced analytics are outputted into \textit{summaries}, stored permanently within the \textit{Data Storage} service.

The \textit{Process Engine} is composed by processing topologies for every type of sensor. Each topology has the ability to be easily modified in order to accommodate aggregation operations. The processing topology is comprised of \textit{Aggregator} steps that are responsible for the analysis of the data based on specific time intervals. Each processing topology aggregates data for specific time intervals (see Figure \ref{fig-sparks-chain}). An \textit{aggregator} stores all the events of the time interval and for each new incoming event it processes (using functions such as min/max/mean) all the stored values of this interval. After this process, it updates the existing interval value which is used by the next process level.

\begin{figure}[!t]
\centering
\includegraphics[width=0.37\columnwidth]{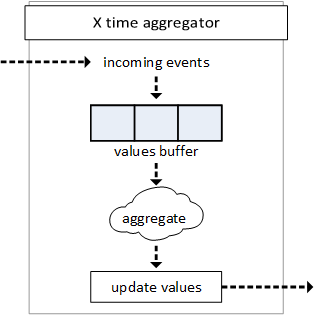}
\caption{\label{fig-sparks-aggrgator} Aggregator module}
\end{figure}

Events which enter the continuous computation engine are processed consecutively. First, the processing topology performs aggregation operations on the streaming data, i.e., for a temperature sensor, the engine will calculate the average values of the 5 minute interval (see Figure \ref{fig-sparks-aggrgator}) and it will store it to memory and disk for further process (when the processing topology receives more than one events for the same 5 minute interval, it calculates the average of those events). Every consecutive 5 minute interval aggregate is kept in memory (topology keeps $48$ interval values $k$ for 5min, hour, day, month intervals for each device) / stored by the \textit{Data Storage} service. Next step is to update the hour intervals. The processing topology updates the 5min-intervals inside the buffer of the hour processor and stores the average of those 5min-intervals. 

The process is similar for the daily processor, but the topology also stores the max/min of the day (based on hour intervals). Same for monthly and yearly processors. For power consumption sensors, the scheme (topologies inside Storm) is the same with the difference that the topology calculates and stores power consumption. Aggregators are used to perform aggregation operations on input streaming data.  The topologies use aggregation for \textit{Power Consumption calculation} (calculate the power consumption of the stream values), \textit{Sum calculation} (summarize the streaming values), \textit{Average Calculation} (calculate the average of the streaming values).

An important characteristic for evaluating the performance for the system is the load of data the system is able process at any given time. Having in mind the current setup, the fleet of buildings in the system produces an average of $25$ measurements per second. The current data processing topology runs on a single core virtual machine (on a Intel i5-3340 host) with $4GB$ of RAM. With this configuration and setup the system is capable of processing up to $500$ measurements per second. To increase the number of measurements our system can support there are two different options:
\begin{itemize}
 \item Increase the computing power of the virtual machine, by assigning it to a more powerful host or giving access to more resources from the host machine.
 \item Deploy a second instance of the processing topology that is capable to consume the same number of measurements to reach the required data processing rates.
\end{itemize}

Based on the nature of sensor deployed in our system input data require $3$ different types of aggregation: (1) averaging for sensors like temperature or relative humidity, (2) total for sensors like rain height levels, (3) power consumption estimating based on the electrical current values received from the installation. Each type of processing requires a different type of aggregation processing and as a result has a different average execution latency, presented in Table~\ref{table:processing}. 

\begin{table}
\centering
 \caption{Execution Latency statistics for the 3 different aggregation types used in our system.\label{table:processing}}
 \begin{tabular}{|l|c|c|}
  \hline
  Aggregation Type & Execution Latency (ms) & Measurements (\%) \\ \hline
  Average &  0.608 & 86.4 \\ \hline
  Total & 0.799  & 0.9 \\ \hline
  Power Consumption & 0.329 & 12.7 \\ \hline
 \end{tabular}
\end{table}

\subsection{Data Access and Acquisition}

The platform provides a unified API for retrieving data from multiple sites and multiple hardware platforms transparently. Each hardware device integrated to our platform is mapped to a
\textit{resource}. Resources are self-described Entities and are also software/hardware agnostic. The Data API acts as a wrapper function and hides much of the lower-level plumbing of hardware specific API calls for querying and retrieving data and provides a common API for retrieving historical or real-time data from resources in a transparent manner.

To facilitate integration between the existing hardware and software technologies, the exchange of the information occurs through \textit{API Mappers}. The API Mapper acts as a translation proxy for data acquisition and it is responsible for polling the devices infrastructure through proprietary APIs and translating the received measurements in a ready to process form for the platform. In general, the API Mapper transforms data to and from the API. The data input type can be, based on each device capabilities, poll based and/or push-based. In more details, the API Mapper is capable to receive data from the IoT devices, but also to send messages/commands to the devices. Furthermore, according to the system design, the API Mappers introduce scalability and modularity in the platform. The solution offers two separate types of API Mappers for integrating with external services and to retrieve IoT sensor data: a) Polling API Mapper, and b) Message Bus API Mapper. Both solutions will be used in order to integrate with data originating from IoT installation.

The first solution (Polling API Mapper) is based on polling. A usage example is the following: weather stations are installed in a subset of available school buildings. Data produced by  such stations are accessible through the SynField application that provides historical information through a RESTful API provided by the SynField back-end. In  order to integrate them in our platform, a SynField API Mapper was implemented for the SynField API based on the Polling API Mapper. The SynField API is polled every 5 minutes for updated data. When new data are found, they are formatted to the internal format of the platform and forwarded to the Processing/Analytics engine, for processing and analysis using the AMQP protocol. The data are then processed and can be accessed from the Data API. Similar implementations based on the Polling API Mapper can be used to integrate IoT devices provided by third parties and existing BMSs  in school buildings. 

The second solution is used when a pub/sub solution exists in the external service to be integrated. In this case, the external service is capable of publishing the IoT data (generated or gathered) to an MQTT endpoint. The API Mapper is then able to receive new measurements asynchronously and format them to the internal format of our platform. The data are then forwarded to the \gaia Processing/Analytics engine for processing and analysis using the AMQP protocol. The data are then processed and can be accessed from the \gaia Data API. Messages inside the MQTT broker can be transferred in multiple formats ranging from plain text to any open or proprietary protocol. In our case messages are transmitted in plain text following a simple format: the topic of the message refers to the device and sensor that generated the message while the actual payload represents the value generated. For example, if a sensor with a hardware (MAC) address \texttt{124B00061ED466} publishes a temperature value of $20$ degrees, the topic is \texttt{124B00061ED466/temperature} and the message \texttt{20}. All sensors forward their measurements periodically (every 30 seconds) or on events (i.e., when motion is detected) and the API mapper receives them and forwards the to the processing engine.

\begin{figure}
  \centering
  \includegraphics[width=\columnwidth]{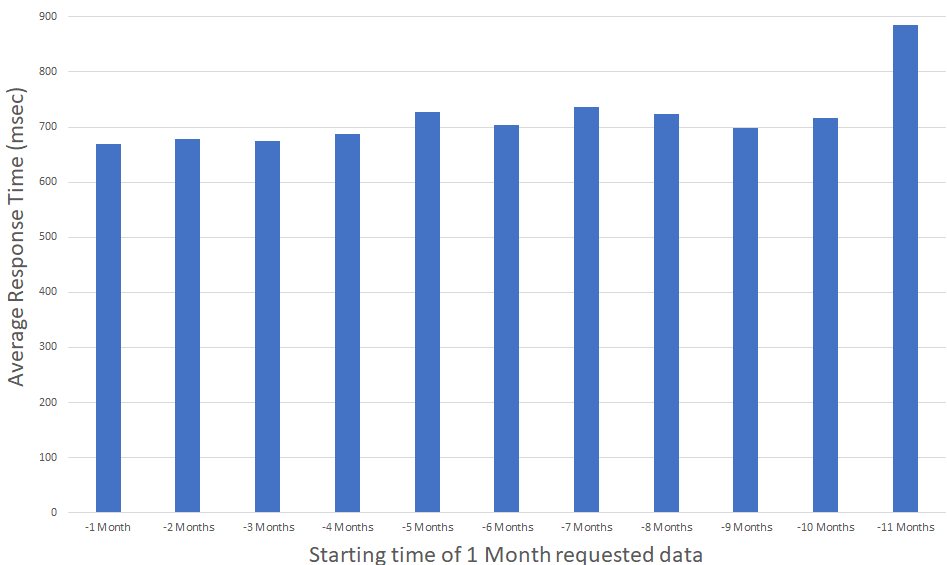}
  \caption{Average Response Time for accessing $1$ month data for the past year (daily aggregated values).}
  \label{fig:eval:DataAccess1}
\end{figure}

\begin{figure}
  \centering
  \includegraphics[width=\columnwidth]{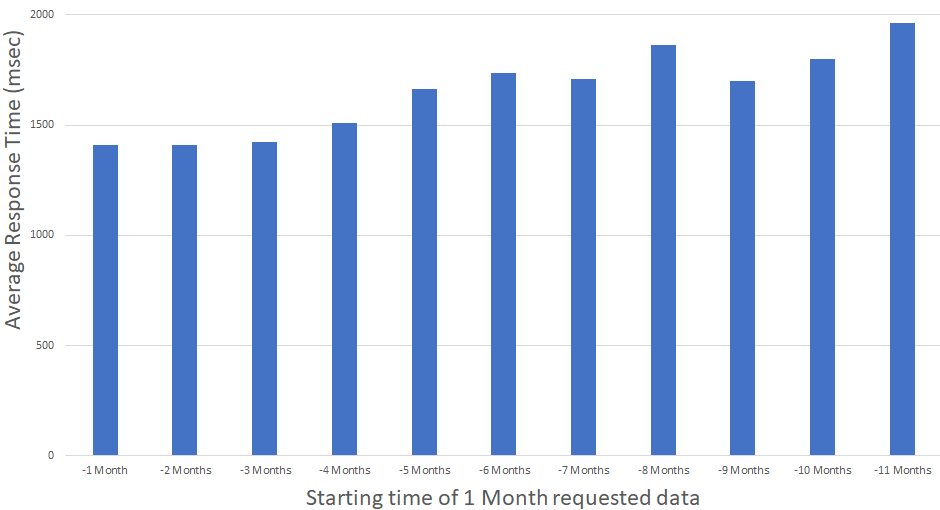}
  \caption{Average Response Time for accessing $1$ month data for the past year (hourly aggregated values).}
  \label{fig:eval:DataAccess2}
\end{figure}

Accessing historical data is crucial for building monitoring applications. Users tend to search for and compare historical data from different timespans and areas of the buildings. In such use cases, it is important that an IoT service is capable of providing these data without delays independently of the targeted time interval. As dicussed in \cite{Card:1991:IVI:108844.108874}, application response times larger than $10$ seconds tend to make users loose their attention in the given task while a $1$ second response time is considered the limit for users that are freely navigating an application without waiting for the application's response. In that context, when presenting power consumption statistics for example over the past year it is important to be able to retrieve and present the stored values in under 1 second independently of the requested interval (latest values versus older values. In Figures~\ref{fig:eval:DataAccess1} and \ref{fig:eval:DataAccess2},  we present average retrieve times for accessing historical data of $1$ month duration for the past $12$ months, observing minimal differences in the access times independent of the period requested. 

\begin{figure}
  \centering
  \includegraphics[width=\columnwidth]{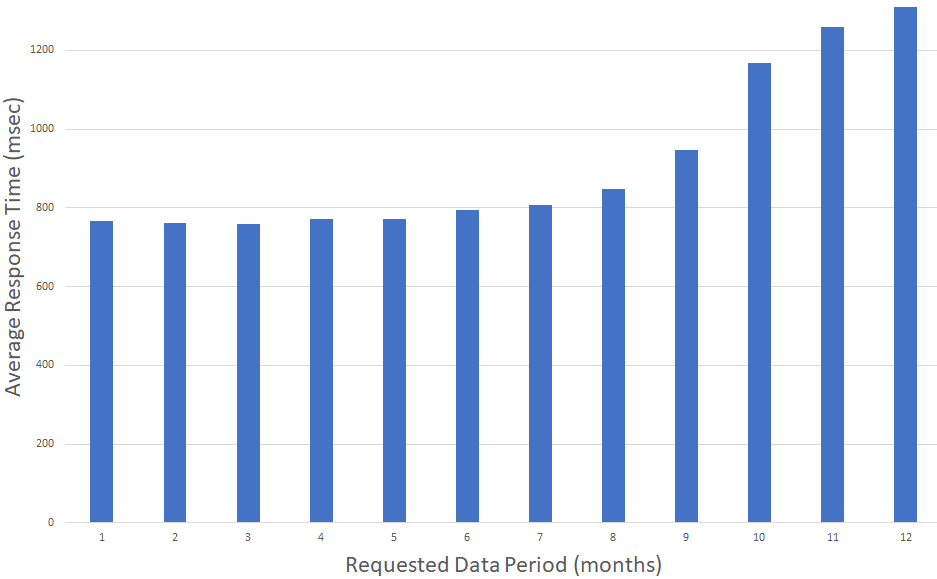}
  \caption{Average Response Time for variable time periods ranging from $1$ to $12$ months.}
  \label{fig:eval:DataAccess3}
\end{figure}

Note that the data available from the graphs that the response time of our service is independent of the actual time interval while it is actually dependent on the amount of data requested. This is more clear in Figure~\ref{fig:eval:DataAccess3} where it is observed that response times tend to increase as the response times increase when we reach time periods of more than $9$ months of data.

\section{Using IoT-generated Big Data in Educational Buildings}
\label{sec:data}
Ensuring the energy efficiency and sustainable operation of school buildings is an extremely complex process since each building has vastly different characteristics in terms of size, age, location, construction, thermal behavior and user communities. IoT platforms can provide quantitative evidence to evaluate and improve organizational and managerial measures. Some aspects of the business value and benefits of analyzing the data collected from the IoT platform are presented in this section.  

\subsection{High-level IoT Data Analysis}

The IoT deployment of the \gaia platform is continuously expanding to include additional school buildings. The collected data was analyzed to determine the variety of the sensors supported and their points of sensing (POS), the velocity of data arriving at the cloud infrastructure, as well as the variability of the data collected. 

\begin{figure}
\centering
\includegraphics[width=0.9\columnwidth]{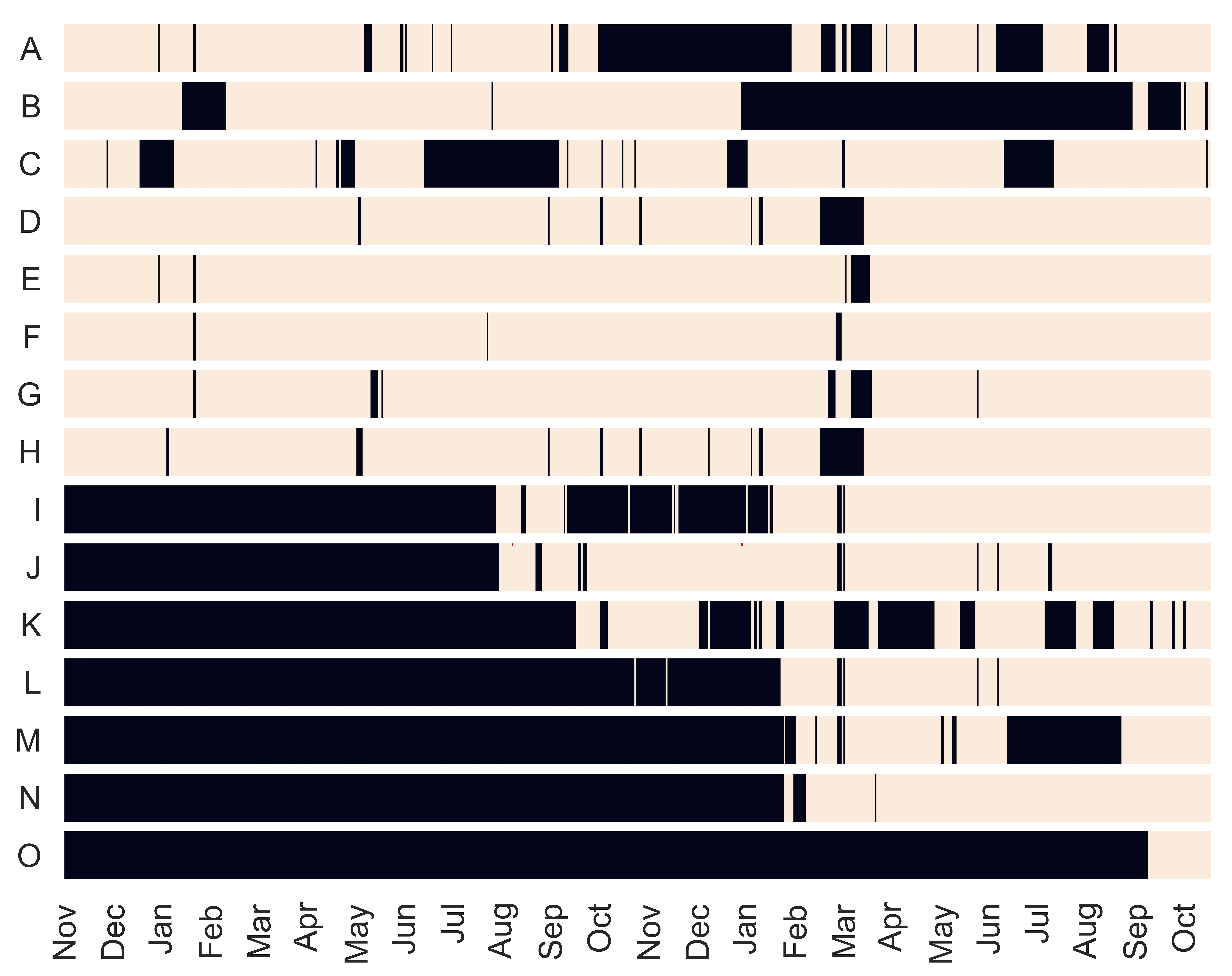}
\caption{\label{fig-data-1} Data availability per School Building}
\end{figure}

During this initial high-level analysis it became apparent that the data collected from the IoT devices was not always delivered properly to the cloud. As a first step towards understanding the
availability of measurements, Fig.~\ref{fig-data-1} is included that depicts the availability of measurements on a daily basis, organized based on the site of deployment. A single dark point signifies missing data for the specific on a specific day, while a white part signifies complete availability (i.e., according to the sensing rate of the specific sensor). This visualization shows the stability of
the IoT deployment. In almost all deployments there are values missing almost on a daily level. Essentially these measurements are missing either because they were never reported to the cloud infrastructure or due to a failure occurring while they were processed and stored by the cloud services. In the first case, network failures occur either due to a packet transmission error at the wireless network level (i.e., \ieeeZ  or \wifi), or due to a transmission error while an intermediate gateway transmitted the data to the cloud infrastructure over the Internet. Since the access to the
measurements are done through the \gaia platform API, the reason for the missing information is completely unknown. However, as it will become evident in the following sections, specific data mining techniques can be used to overcome the problem of missing values.

\begin{table}
\centering
\caption{Data availability per School Building}
\label{table:table-data}
\begin{tabular}{|c|c|c|c|r|c|}
\hline
Site & POS & Sensors & Start time & Outages & Outliers \\\hline
A  & 8 & 43 & 2015-OCT & 17.78\% & 2.67\% \\
B & 9 & 56 & 2015-OCT & 28.26\% & 2.84\% \\
C & 6 & 32 & 2015-OCT & 13.63\% & 2.82\% \\
D & 9 & 50 & 2015-OCT & 2.33\% & 1.45\% \\
E & 7 & 43 & 2015-OCT & 1.19\% & 1.63\% \\
F & 11 & 54 & 2015-OCT & 0.96\% & 3.36\% \\
G & 7 & 45 & 2015-OCT & 1.69\% & 1.33\% \\
H & 6 & 27 & 2015-OCT & 6.01\% & 1.31\% \\
I & 7 & 36 & 2016-SEP & 12.23\% & 1.68\% \\
J & 34 & 103 & 2016-APR & 3.97\% & 3.19\% \\
K & 5 & 26 & 2016-SEP & 15.87\% & 1.7\% \\
L & 5 & 109 & 2016-OCT & 36.43\% & 1.22\% \\
M & 5 & 24 & 2017-FEB & 23.09\% & 2.67\% \\
N & 12 & 55 & 2017-FEB & 1.31\% & 5.09\% \\
O & 4 & 22 & 2017-SEP & 0\% & 13.22\% \\ \hline
\end{tabular}
\end{table}

A second step for the high-level analysis is to examine the actual
values received from the IoT devices. It is very common in the relevant
literature to deploy relatively low-cost devices that produce
low-quality measurements or are not properly calibrated. For this
reason we examined the values to identify possible outliers, that is
observation points that are distant from the historic values. Such
observations may be due to transient errors occurring on the sensing
equipment and should be excluded from the data set. The identification
of outliers is based on the \textit{interquartile range} ($IQR$) using
the upper and lower quartiles $Q_3$ (75th percentile) and $Q_1$ (25th
percentile). The lower boundary is set to $Q_1 - 3 \times IQR$ and
the upper bounder is set to $Q_3 + 3 \times IQR$ where $IQR = Q_3 -
Q_1$.
The values are examined per sensor/site basis using a timed-window of
size $W$. If the a value is outside the boundaries $\left[Q_1 - 3 \times
IQR, Q_3 + 3 \times IQR\right]$ it is flagged as an outlier. In the
following sections we replace it with the minimum or the maximum value
observed during the time window $W$. After examining the values characterized
as outliers two distinct cases were identified: (a) $0$ values which
were clearly sensor error rather than natural events (e.g. humidity of $0\%$,
temperature dropping from $\sim 20$ to $0$), and (b) drastic changes
of power consumption (i.e., spikes or fast drops) that could not be justified
by the daily school activities.

\begin{table}
\centering
\caption{Data availability per Sensor Type}
\label{table:table-device}
\begin{tabular}{|l|c|c|r|r|}
\hline
Name 		& POS 	& Sensors 	& Inactive 	& Outlier \\ \hline
Environmental	& 101 	& 505 		& 14.62\%	& 7.76\% \\ 
Atmospheric  	& 7 	& 56 		& 19.56\% 	& 6.29\% \\
Weather 	& 7 	& 28		& 20.25\% 	& 0.95\% \\
Power  		& 20 	& 56 		& 12.55\% 	& 4.17\% \\ \hline
\end{tabular}
\end{table}

In Table~\ref{table:table-data}, the different school sites are
summarized indicating the time when they were incorporated in the \gaia
platform. For each school building, the number of points of sensing (POS)
are listed along with the total number of sensors deployed. The table
reports the percentage of outages recorded for the particular site
(reflecting the periods during which no measurements were received
from it, as a percentage of the period from when it was first
incorporated to the platform), the total number of 
measurements received from this site, along with the percentage of values that have been
identified as outliers. In order to avoid issues related to the
confidentiality of data, the names of the school buildings have
been omitted.

\begin{figure}
\centering
\includegraphics[width=0.9\columnwidth]{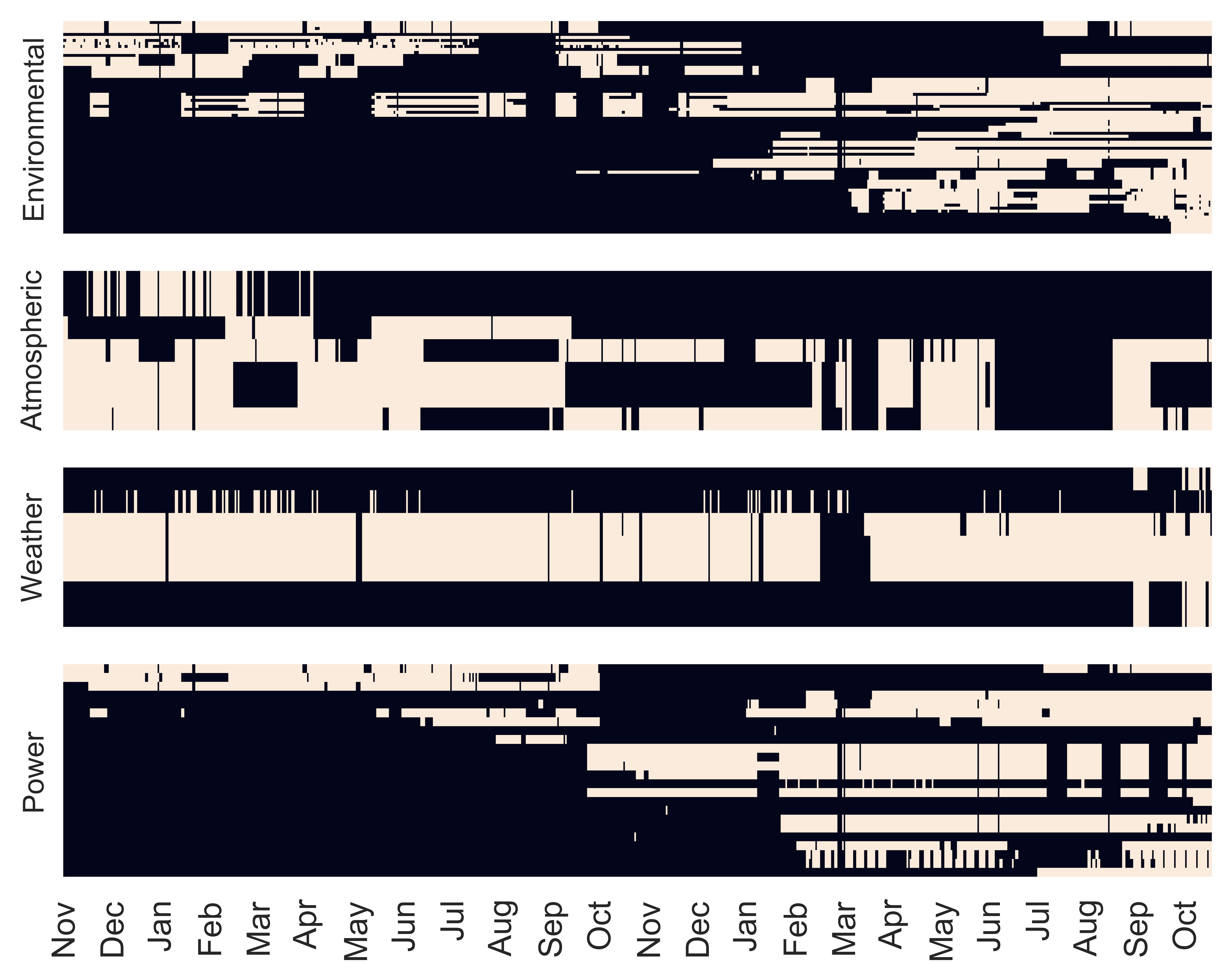}
\caption{\label{fig-data-2} Data availability per Sensor Type}
\end{figure}

The analysis reveals that certain buildings experience data outages very
often. At a second level, the same analysis of data is repeated based on the type of sensor. For each device category, the percentage of outages and the percentage of
outliers observed are reported in Table~\ref{table:table-device}. 
In Fig.~\ref{fig-data-2}, the availability of data is depicted on a daily basis for each sensor separately organized by
sensor category. Based on this, one observes that all sensors
experience periodic loss of data. This may be justified by the wireless
networking technology used to interconnect the sensors located in the classrooms, as
reported in~\cite{s17102296}. Apparently the low-power, lossy nature of 
the networking technologies used results to a significant data loss.

\begin{figure}[!b]
\centering
\includegraphics[width=.9\columnwidth]{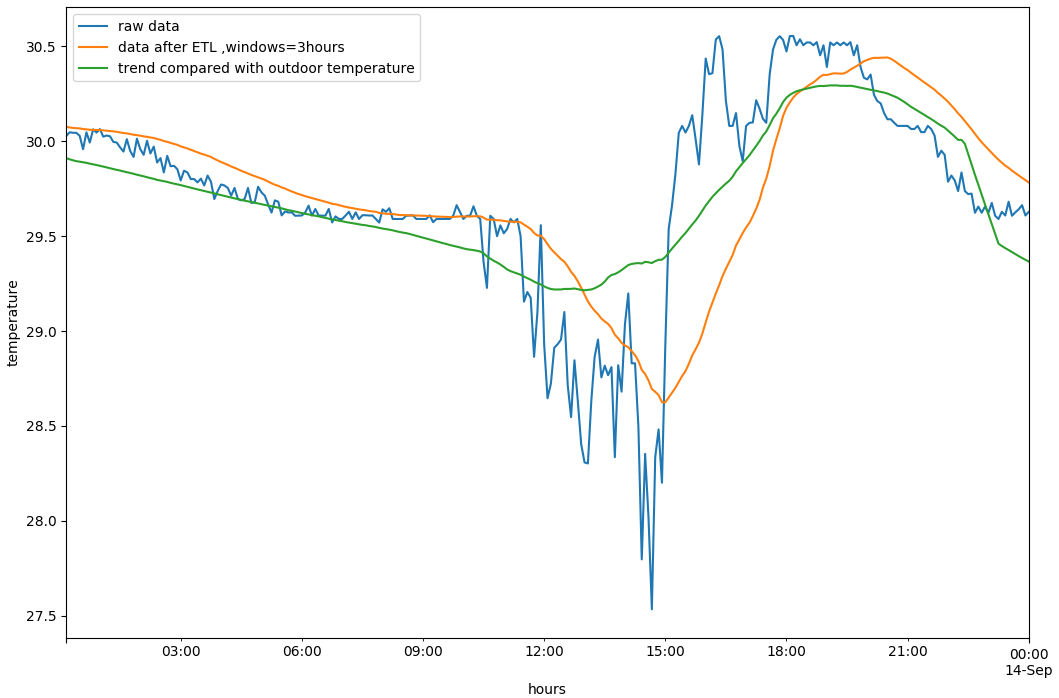}
\caption{\label{fig:etl} Temperature sensor time series processing}
\end{figure}

To overcome the fact that the IoT deployment (1) experiences outages on a regular basis, and (2) at a significantly lower rate, sensors report values characterized as
outliers, a moving window average technique is used. The moving window (1) smooths out short-term fluctuations for the case of outliers and (2) fills-in missing values using a simple local algorithm that introduces historic values to fill in the missing data for the specific time period. The moving window average also helps to highlight longer-term trends on the sensor values. Fig.~\ref{fig:etl} depicts an example of the analysis conducted over a specific temperature sensor located in a classroom.

\subsection{Thermal Comfort of Classrooms}

\begin{figure}[!t]
\centering
\includegraphics[width=\columnwidth]{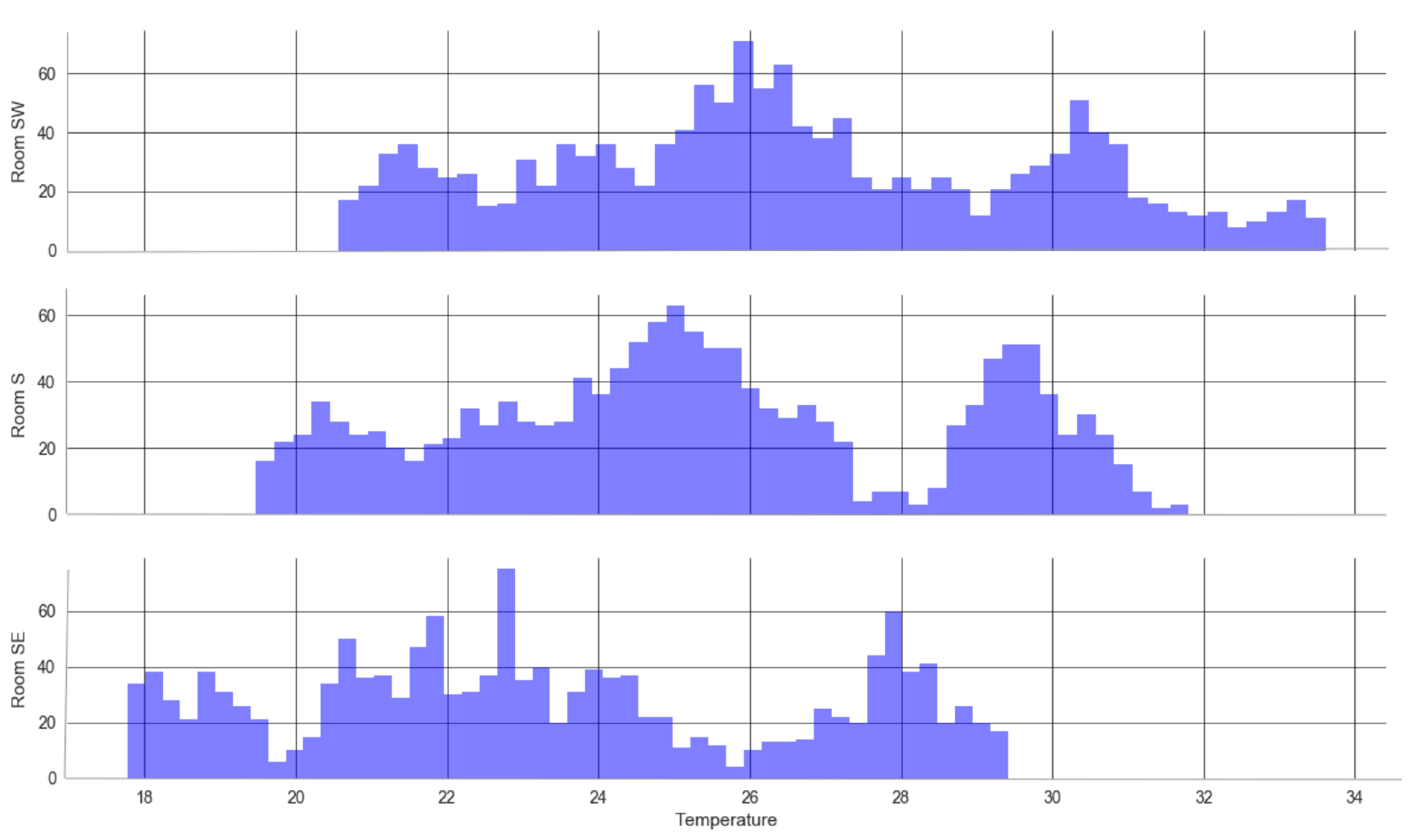}
\caption{\label{fig:histogram} Indoor temperature histogram for 3 classrooms during Sep/17 to Oct/17}
\end{figure}

Examining the classrooms' temperature is a measure of understanding the conditions under which students and teachers operate. Hot, stuffy
rooms---and cold, drafty ones---reduce attention span and limit productivity. In Fig.~\ref{fig:histogram} a histogram is provided for the indoor
temperature of 3 classrooms (facing south, south-west and south-east) examined during a period of 2 months. Lower temperatures are observed in the room facing South-East, in contrast to the other two. Evaluating the indoor conditions requires considering also other factors related to the environment, such as humidity (e.g., excessively high humidity levels contribute to
mold and mildew), as well as what students are wearing (depending on the period of the year).

A common approach to examine the indoor conditions of the classrooms is
in terms of the \emph{thermal comfort}. The ANSI/ASHRAE
Standard 55, Thermal Environmental Conditions for Human Occupancy~\cite{ashrae} 
is defined to specify the
combinations of indoor thermal environmental factors and personal
factors that will produce thermal environmental conditions acceptable to
the majority of the occupants of an area. Thermal comfort is
primarily a function of the temperature and relative humidity in a room,
but many other factors affect it, such as the airspeed and the
temperature of the surrounding surfaces.
Furthermore, thermal comfort is strongly influenced by how a specific room
is designed (e.g., amount of heat the walls and roof gain or
lose, amount of sunlight the windows let in, whether the windows can
be opened or not) and the effectiveness of the HVAC system. 

The analysis presented here is done
based only on the quantitative information provided by the IoT
deployment of the \gaia project, we applied a simple approach that is
able to provide an estimate under the lack of all the necessary data.
For this reason the CBE Thermal Comfort Tool for ASHRAE-55\cite{cbe} 
was used to evaluate the thermal comfort of classrooms. 
Under the \emph{adaptive method} provided, thermal comfort is computed as a function of the indoor
temperature, the outdoor temperature along with the outside air wind
speed. The outside conditions (temperature, airspeed) were acquired by
the \emph{WeatherMap} service that also provides historical data. For
each different classroom, the thermal comfort is computed for each
different hour during the operation of the school (from 08:30 to 16:30). 
In the sequel, the individual values are averaged over each different day. 
A classroom with a daily comfort
of $1.0$ signifies that during all hours the conditions where within the
comfort zone defined by this particular formula, while a daily comfort
of $0.0$ signifies that during all hours the conditions were outside the
comfort zone. 

\begin{figure}[!t]
\centering
\includegraphics[width=0.85\columnwidth]{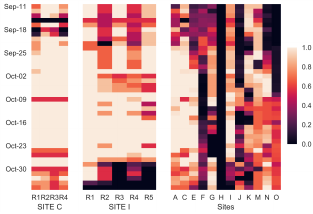}
\caption{\label{fig:comfort} Site thermal comfort during Sep/17 to Oct/17}
\end{figure}

On the right side of Fig.~\ref{fig:comfort}, a summary of all the sites
participating in the \gaia platform is provided for a period of
2 months (from Sep/2017 to Oct/2017). On the left side, the individual thermal
comfort of the classrooms of site $C$ is compared with those of site $I$.
Site $C$ achieved the highest comfort in contrast to site $I$ that achieved the
lowest. One reason for the difference is the actual location of the school,
site $C$ being the school located on the southern point in contrast to site $I$
which is one of the northern points. Apart from the external weather conditions, other reasons affect thermal comfort, e.g., such as the construction materials.
In the following section, data mining techniques are applied in order to identify in more
details classrooms with poor performance or user-related activities that may also affect thermal
comfort.

\subsection{Classroom Thermal Performance}


Several factors affect the internal classroom temperature
ranging from the local weather conditions to the orientation of the
room, the construction materials used (e.g., the insulation, windows)
and also the position of the radiators within the classrooms. The \gaia
platform includes school buildings located in different climatic zones,
constructed in different years ranging from 1950 to 2000, using diverse
materials and with different heating and ventilation systems.
Unfortunately, such information is not available in an open format so
that they can be incorporated via data integration techniques. In this
section, data mining techniques are used to derive information regarding
the thermal efficiency of the classrooms using the quantitative data collected from
the IoT deployment. 

The goal of the analysis is to identify classrooms with poor thermal performance. 
One of the leading factors affecting temperature in classrooms is orientation.
During a sunny day, at mid-day the classrooms experience the highest temperatures.
We can also see (Fig.~\ref{fig:histogram}) that classrooms facing south-west are exposed to the sun for longer periods, thus maintaining higher temperatures for longer than classrooms with a different orientation. In order to factor in the potential contribution of the sun, the time-series of the temperature sensors are examined in correlation with the external temperature, the cloud coverage, as well as the orientation of the classrooms.

A second important factor that affects the internal temperature of classrooms is the
daily activities of the students and teachers. Opening and closing the door, the windows and the window blinds have an immediate effect on the temperature. For example, when a window is open there
is a temperature drop of about $2^oC$. In order to overcome these effects, the performance of the
classrooms is examined only during weekends when there are no school activities.

\begin{figure}
\centering
\includegraphics[width=\columnwidth]{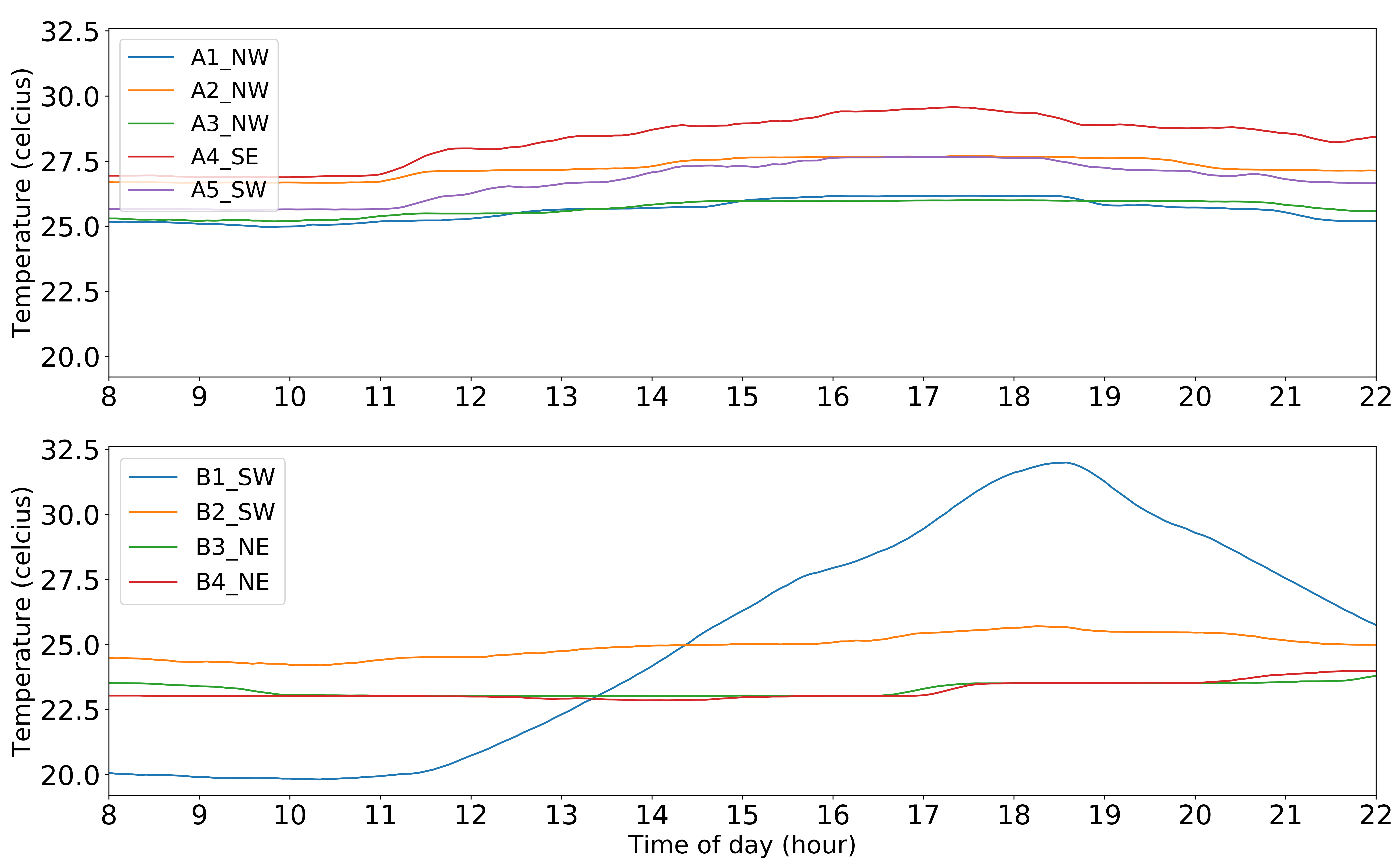}
\caption{\label{fig:weekend} Classroom temperature during 30/Sep (Saturday)}
\end{figure}

Given the above considerations, we examine temperature in each room to identify
poorly performing classrooms. In Fig.~\ref{fig:weekend}, 2 specific performance issues regarding two schools located in the same city are depicted. The first issue is related to the bottom figure, 
where room R1 achieves very poor performance with temperature starting very low at $20^oC$ 
and increases up to $32^oC$ within 8 hours. The second issue is related with the top figure, 
where the south-west facing classroom (R5) and the south-east facing classroom (R4) have an increase of 2 degrees during the day, while all the other rooms are not affected. Even the south-west
facing room R2 of the bottom figure does not have such an increase during the day. 
After contacting the school building managers it was reported that (a) room R1 (bottom school)
is located outside the main building, within a prefab room with poor insulation, and
(b) rooms of top school have no window blinds installed, in contrast to the bottom school where window blinds
are installed in all rooms. These are just examples of the results of the analysis conducted. 
We expect that such an analysis can provide strong evidence on how to improve the performance of schools.

\begin{figure}
\centering
\includegraphics[width=\columnwidth]{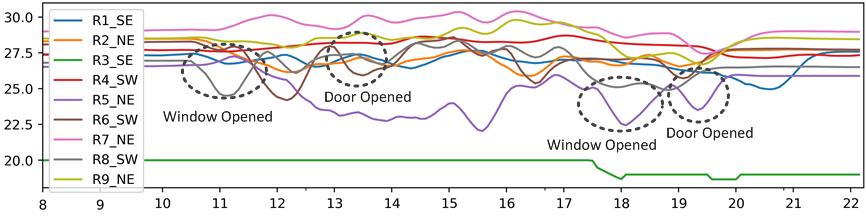}
\caption{\label{fig:events} Activity in classrooms and effect on temperature}
\end{figure}

A second goal of the analysis is to understand how certain user activities affect the indoor quality of
classrooms. The performance of the rooms during weekends when there is no user activity is compared with
days of similar external conditions (temperature, cloud coverage) when students are present. 
In Fig.~\ref{fig:events} the temperature of the classrooms of a specific school is shown.
The figure includes annotations for four instances demonstrating the impact of opening/closing the
classroom door and windows. It is evident that these 4 specific events have an immediate impact on the temperature, with the window-related events having a greater impact. Automatic identification of such events in real-time may be important for teachers to understand when to open the windows and doors for fresh air to circulate in the
classroom. Automatic notifications can be sent when the thermal comfort of a classroom is outside a certain range in order to reduce the temperature (e.g., by opening door/windows).

\section{Conclusions}
\label{sec:conclusions}
\label{sec:conclusion}
In this chapter, the use of IoT sensors in educational environments was discussed. An education-focused real-world IoT deployment in schools in Europe can help promote sustainability and energy awareness. The chapter presents how installing IoT sensors inside school environments and making them available to students and other stakeholders of the educational community, can aid in pursuing goals with respect to environmental education and science curricula alike. Large datasets, in particular, are being acquired from classrooms inside school buildings, subject to subsequent data analysis. Results of such analysis has, in turn, prompted  human actions onto the actual environment, in order to achieve a more comfortable environment (i.e., indoor comfort levels) in a more energy-efficient manner. By using this infrastructure and the data it produces, we can build tools that better reflect the everyday reality inside school buildings, providing a more meaningful feedback. An interesting future research direction for the IoT infrastructure is to cater for monitoring additional environmental parameters apart from energy consumption, for example in order to collect data to further promote sustainability awareness and behavioral change, and such big data sets be utilized in educational scenaria.

\section*{Acknowledgements}
This work has been supported by the European Commission and EASME, under H2020 and contract number 696029. This document reflects only the authors’ view and the EC and EASME are not responsible for any use that may be made of the information it contains.

\bibliographystyle{abbrv}
\bibliography{chapter-gaia}

\end{document}